# Double Magnetic Transition in $Pr_{0.5}Sr_{0.5}CoO_3$

R. Mahendiran and P. Schiffer

Department of Physics and Materials Research Institute, Pennsylvania State University,

University Park, PA 16802

We report studies of polycrystalline samples of the metallic ferromagnet $Pr_{0.5}Sr_{0.5}CoO_3$ through measurements of the magnetization, a.c. magnetic susceptibility, resistivity, and specific heat. We find an unusual anomaly around $T_A$ = 120 K, much below the ferromagnetic transition ($T_C$ = 226±2K). The anomaly is manifested in field cooled magnetization as a downward step in low fields (H ≤ 0.01 T) but is transformed into an upward step for H ≥ 0.05 T. The anomaly cannot be easily attributed to antiferromagnetic ordering, but may correspond to a second ferromagnetic transition or an alteration of the ferromagnetic state associated with orbital ordering.

The past few years have witnessed a renaissance of interest in mixed valent transition metal oxides of the type $R_{1-x}A_xMO_3$ (R = $La^{3+}$, $Pr^{3+}$, etc., A = $Ca^{2+}$, $Sr^{2+}$, etc., M = Mn, Co),[1,2] but the origins of ferromagnetism and magnetoresistance in cobaltates seems to be fundamentally different from that of the manganites.[3] The phase diagram of cobaltates is also much simpler than manganites which exhibit exotic varieties of antiferromagnetic phases with different doping level (x). The widely studied $La_{1-x}Sr_xCoO_3$ series shows spin glass ($0.05 \leq x \leq 0.2$) and cluster glass ($0.3 \leq x \leq 0.5$) behavior and $SrCoO_3$ is a long range ferromagnet.[4] The cluster glass phase has long range ferromagnetic order but with a possible coexistence of superparamagnetic clusters.[5] The use of cobaltates in ferroelectric thin film capacitors,[6] solid oxide fuel cells,[7] possible applications as magnetostrictive actuator[8] and thermoelectric element[9] and more importantly their distinct physical properties with respect to manganites are compelling reasons to investigate them in detail. We have investigated the magnetic, electrical, and thermal properties of a Pr-based ferromagnetic metallic cobaltate, and we find evidence for a double magnetic transition unlike that observed in previous studies of the cobaltates.

Polycrystalline $Pr_{0.5}Sr_{0.5}CoO_3$ was prepared by a sol-gel process previously used[8] in the synthesis of $La_{0.5}Sr_{0.5}CoO_3$, and the qualitative features presented below were confirmed in samples prepared by a standard ceramic synthesis. The oxygen stoichiometry was found to be 2.97±0.02 from iodometric titration and the room temperature structure was found by X-ray diffraction to be monoclinic ($P2_1/m$) in accordance with an earlier report.[10] The d.c. magnetization (M) was measured using Quantum Design SQUID magnetometers (MPMS), and a.c. susceptibility ($\chi$), resistivity, and specific heat (C) were measured with the Quantum Design Physical Property Measuring System (PPMS). The temperature dependent magnetization, M(T), was recorded in three modes. In the ZFC mode, the sample was first cooled to T = 5 K in zero field and data were taken while warming after establishing a magnetic field at 5 K. In field-cooled-cooling (FCC) and field-cooled-warming (FCW) modes, the field was applied at 300 K, and data were taken during cooling and warming, respectively in a field. We measured $\chi(T)$ in different d.c. bias fields ($H_{dc}$) in FCW mode with $H_{ac}$ = 10 Oe rms.

Figure 1 shows the temperature dependence of the inverse susceptibility (H/M) at H = 10 mT while warming the sample from 5 K after zero field cooling. The data in the temperature range 255 K -370 K fit a Curie-Weiss law, M = C/(T- $\Theta$) with $\Theta$ = 241.5 K and C = 1.84 emu/mole K. The $Co^{3+}$ and $Co^{4+}$ ions can be in either the low spin (LS) state, the intermediate spin (IS) state or the high spin (HS) state due to the closeness of the crystal field and exchange energies.[11] In the ferromagnetic metallic composition of $La_{1-x}Sr_xCoO_3$ (0.3 < x < 0.5), the $Co^{3+}$ and $Co^{4+}$ ions are believed to be in the IS state ($t_{2g}^5 e_g^1$, S= 1) and the LS state ($t_{2g}^5 e_g^0$, S = 1/2) respectively.[11] The estimated effective paramagnetic moment $P_{eff}$ = 3.84$\mu_B$ from the experimental C value of our sample is higher than $P_{eff}$ = 3.45$\mu_B$ calculated with 50 % IS $Co^{3+}$ ($P_{eff}$ = 2.84 $\mu_B$), 50 % LS $Co^{4+}$ ($P_{eff}$ = 1.73$\mu_B$) and 50 % $Pr^{3+}$ ($P_{eff}$ = 3.58 $\mu_B$).[12] The observed and calculated $P_{eff}$ values do match, however, if half of the LS $Co^{4+}$ are in the IS state ($P_{eff}$ =3.87 $\mu_B$). Such a combination of IS $Co^{3+}$, IS $Co^{4+}$ and LS $Co^{4+}$ is consistent with the maximum value of the magnetization shown in the inset of figure 1 ( M = 1.87 $\mu_B$ at T = 5 K and H = 7 T which is close to the saturation magnetic moment of M = 2 $\mu_B$ expected for these spin configurations). While this agreement is reasonable, note that magnetic phase separation[11] or magnetic field induced spin state transitions[8] could affect the relative properties of the different spin-states in the ferromagnetic state, and thus this assignment of the spin states needs to be confirmed by detailed spectroscopic studies.

Fig. 2a shows M(T) at H = 5 mT and 0.01 T. The rapid increase of M around $T_C$ = 226±2K signals the phase transition from a paramagnetic to a ferromagnetic state as expected from neutron scattering studies[10] and from the nature of the M(H) data shown in the inset to figure 1. The magnetization is strongly dependent on magnetic history starting from a temperature just below $T_C$ to the lowest temperature. Although such behavior is known in the related compound $La_{0.5}Sr_{0.5}CoO_3$,[13] there are two unusual features which were not found in the La-based cobaltate. First, there is a clear hysteresis between FCC and FCW curves. Second, there is a downward step around $T_A$ ~100 K in both FCW and FCC curves and a hump in the ZFC data at the same temperature. This anomaly cannot easily be attributed to a transition into an antiferromagnetic state because the M(H) curve at 5 K (discussed below) clearly indicates ferromagnetism, and neutron studies did not observe any evidence for antiferromagnetism.[10] Moreover, the downward step changes into an upward step as H increases to 0.05 T, as seen in figure 2b, and the temperature of the anomaly increases with increasing magnetic field. A similar anomaly can be seen in the relatively high field data of Brinks *et al.*[10] and those of Yoshii and Abe,[4] demonstrating that this lower temperature transition is a robust feature of the material rather than an artifact of our specific sample preparation technique.

Figures 3a and 3b show the temperature dependence of the real part of the a.c. susceptibility, $\chi'(T)$, in different applied d.c. magnetic fields. In low fields, there are two maxima in $\chi'(T)$, one near $T_C$ and one at 70 K. In larger applied d.c. fields, the higher temperature maximum broadens and splits into two peaks, one remaining near $T_C$ and the other moving down in temperature with increasing field (this behavior has also been observed in other ferromagnets[14,15]). The 70 K maximum, the onset of which corresponds to $T_A$, moves to higher temperatures with increasing field and merges with the higher temperature peak for H ~ 1 T. The feature at $T_A$ is also evident in figure 4 where we plot magnetic hysteresis loops at selected temperatures. The coercive field ($H_c$) is rather large (53 mT at 5 K), and, when plotted as a function of temperature (figure 4 inset), there is a clear maximum in $H_c(T)$ around $T_A$. This behavior is very different from that of a conventional ferromagnet in which $H_c$ continuously increases below $T_C$ and suggests that the unusual behavior of the ZFC M(T) is attributable to associated domain effects.

The anomaly at $T_A$ is also manifested as a peak in the zero field specific heat (C) shown in figure 5. The approximate sizes of the peaks in C(T) above the background are 4.7 J/mole-K at $T_C$ and 2.5 J/mole-K at $T_A$ (see insets to figure 5). An estimate of the magnetic entropy associated with these features can be obtained by integrating $\Delta C/T$ after subtracting a smooth background (based on a polynomial fit to C(T) measured above and below the regions of the peaks). The resultant magnetic entropy is $S_{mag}$ = 0.4 JK$^{-1}$mol$^{-1}$ and 0.28 JK$^{-1}$mole$^{-1}$ at $T_A$ and $T_C$ respectively, well below the full spin entropy as expected. By contrast, the zero field resistivity (figure 6) changes its slope around $T_C$ and decreases smoothly without any clear anomaly around $T_A$. Application of a 7 T magnetic field suppresses the change of slope near $T_C$ and results in negative magnetoresistance of about 7 % (similar to La$_{0.7}$Sr$_{0.3}$CoO$_3$[3]) which also does not show an anomaly near $T_A$.

We now discuss possible origins of the anomaly at $T_A$, which appears to be associated with a second ordering transition deep within the ferromagnetic state. Brinks *et al.*[10] observed no clear indication of symmetry breaking structural changes between 300 K and 10 K, although they noted an anomalous change in the unit-cell dimensions with contraction of *c* axis by 0.32 % and expansion of *a* and *b* axes between 10 K and 170 K (possibly associated with an abrupt change in lattice parameters associated with $T_A$). While the downward step in M(T) at low fields suggests an antiferromagnetic transition, the upward step at higher fields discount this possibility, and neutron diffraction studies[10] revealed no evidence of antiferromagnetism at low temperature. The history dependence of M(T) below $T_A$ could indicate a re-entrant spin glass transition, but the upward step observed on cooling at high fields again suggests that this is not the case.

We hypothesize the behavior at $T_A$ indicates either a second ferromagnetic transition or a change in the nature of the ferromagnetic state. This explanation would be consistent with the specific heat and susceptibility peaks at $T_A$, as well as the rise in M(T) upon cooling in large magnetic fields (the drop in M(T) on cooling through $T_A$ at low fields could be associated with the abrupt change in the coercive field noted in Fig. 3). A double ferromagnetic transition would be quite unusual and may be associated with electronic or structural phase separation, i.e. different parts of the sample ordering at different temperatures, or ordering of the Pr moments.[16] Multiple magnetic transitions occurring in a single phase sample has been observed in many of the perovskite manganites[15,17] but we are not aware of previously observations in the cobaltates. An alternative explanation would be a change in the nature of the ferromagnetic coupling associated with orbital ordering among some fraction of the Co ions, e.g. a long range

Jahn-Teller ordering of $e_g$ ($d_{x^2-y^2}$ or $d_{z^2}$) orbitals of the intermediate spin $Co^{3+}$ ion ($t_{2g}^5 e_g^1$). For example, Fauth et al.[18] recently showed evidence for a long range $e_g$-$d_{z^2}$ orbital ordering in $La_{0.5}Ba_{0.5}CoO_3$. Because of the strong magneto-elastic coupling, cooperative orbital ordering within the ferromagnetic state can lead to a change in magnetic domain structure or change in the magnetic anisotropy. Indeed, Liu et al.[19] reported a step in the field-cooled M(T) due to orbital ordering in $La_{1-x}Sr_xMnO_3$ (x = 0.12-0.19), suggesting that this behavior is not uncommon at least in the manganites. In this scenario, the changing character of the feature in M(T) with increasing applied field may be caused by a change in orbital orientation and associated modification in domain structure/spin orientation (which would account for the changes in $H_C$). Another possible explanation of the feature at $T_A$, is that there is a spin state transition of significant fraction of $Co^{3+}$ ions from intermediate ($t_{2g}^5 e_g^1$) to low spin state ($t_{2g}^6 e_g^0$) as T decreases below 120 K, similar to that seen in $LaCoO_3$,[20] but a population of the low spin state is also known to cause insulating behavior contrary to what is observed in our compound.[17]

The above explanations are speculative, and detailed neutron diffraction studies or high resolution synchrotron X-ray diffraction studies would greatly elucidate the nature of the transition at $T_A$. Regardless of the origin of the anomalous behavior at $T_A$, the observed double transition in the magnetization is qualitatively different from the behavior of other cobaltates and deserves further investigation. If the transition at $T_A$ is attributable to orbital ordering, these data will demonstrate a new significance to such ordering in the cobaltates.

Acknowledgments: R. M. thanks Dr. R. Mahesh, Dr. J. Blasco and Dr. A. Maignan for providing different pieces of samples prepared by different chemical routes.

R. M is also thankful to Prof. M. R. Ibarra for his critical comments and B. G. Ueland for assisting in the specific heat measurement. The work was supported by NSF grant DMR–0101318

[20] M. Senaris-Rodriguez and J. B. Goodenough, J. Solid State Chem. **118**, 323 (1995) and references therein.

**FIGURE CAPTIONS**

**Figure 1** Temperature dependence of the inverse susceptibility (H/M) measured while warming with H = 10 mT after zero field cooling to 5 K. The inset shows the field dependence of the magnetization, which indicates the ferromagnetic nature of the low temperature phase.

**Figure 2** Temperature dependence of the dc magnetization of $Pr_{0.5}Sr_{0.5}CoO_3$ under different magnetic history conditions (ZFC-zero field cooled, FCC-field cooled cooling, FCW-field cooled warming). Note the unusual step in FCC magnetization which sets in around $T_A$ = 120 K and changes from downward to upward with increasing field for more than H = 0.05 T.

**Figure 3** The temperature dependence of the real part of the ac susceptibility ($\chi$') in different bias fields (0 ≤ $H_{dc}$ ≤ 0.5 T). Fig. 2(b) shows the data for H ≥ 0.5 T in an enlarged scale. The large peak around $T_C \approx$ 226 K in $H_{dc}$ = 0 T is due to the onset of ferromagnetic transition.

**Figure 4** Main panel: M-H hysteresis loop at few selected temperature. Note that hysteresis at T = 125 K is wider than the one at T = 100 K. Inset: Temperature dependence of the coercive field ($H_C$), which has an anomalous peak near $T_A$.

**Figure 5** Temperature dependence of the specific heat, C(T), in zero magnetic field. The peaks around $T_C \approx$ 226 K and $T_A$ = 120 K correspond to the magnetic transitions

indicated by the magnetization data.  The top and bottom insets show the excess specific heat ($\Delta C$) around $T_C$ and $T_A$ respectively obtained by subtracting a smooth background.

**Figure 6** Temperature dependence of the resistivity ($\rho$) in H = 0 T and H= 7 T. The zero field resistivity exhibits a clear change of slope around $T_C$ = 230 K, but no clear feature is seen around the second magnetic transition ($T_A$ = 120 K).  The temperature dependence of the magnetoresistance (shown by the thin solid line) also shows no feature near $T_A$.

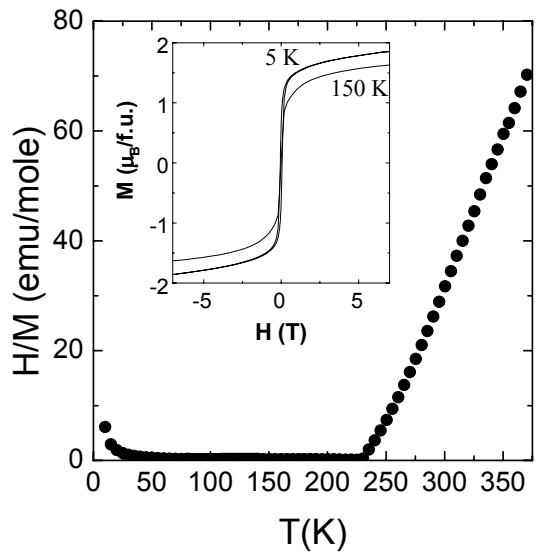

FIG. 1
Mahendiran et al.

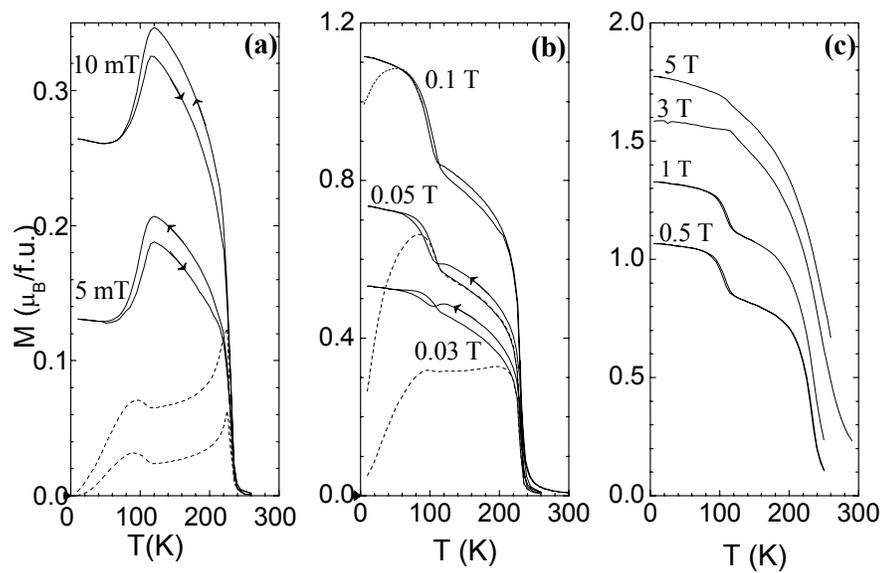

FIG. 2
Mahendiran et al.

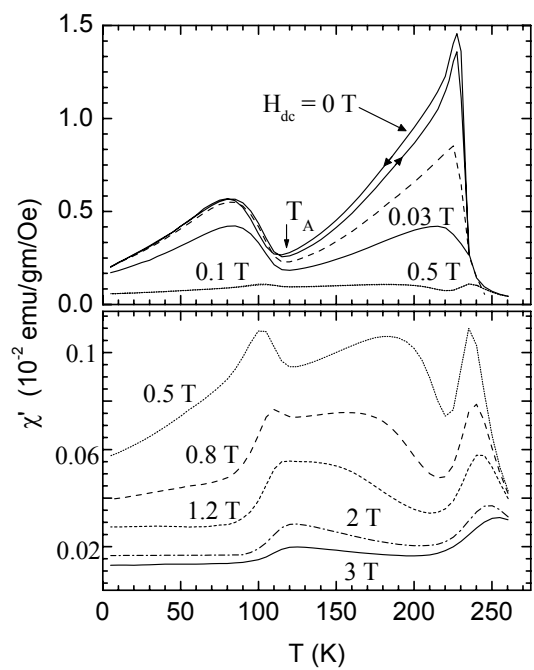

Fig. 3
Mahendiran et al.

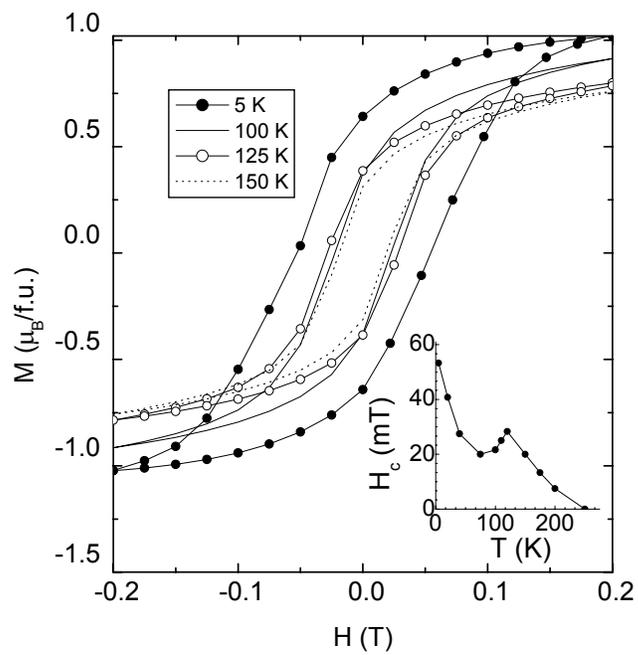

Fig. 4
Mahendiran et al.

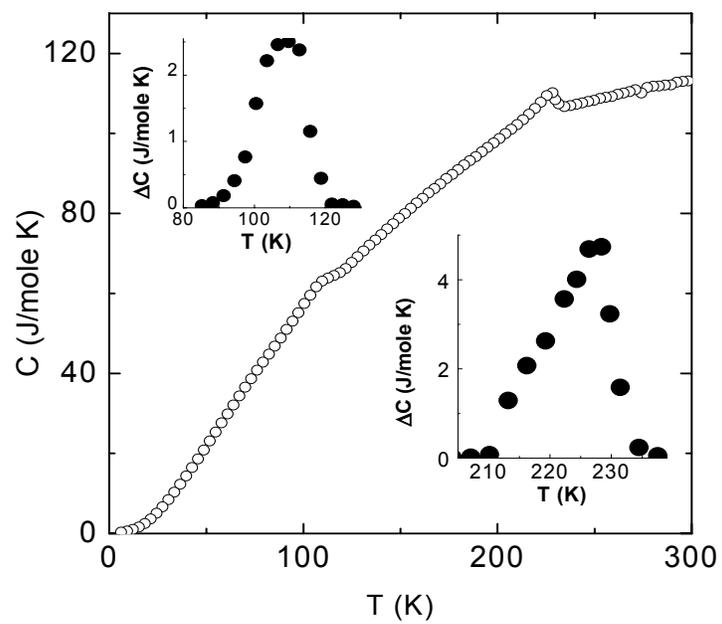

Fig. 5
Mahendiran et al.

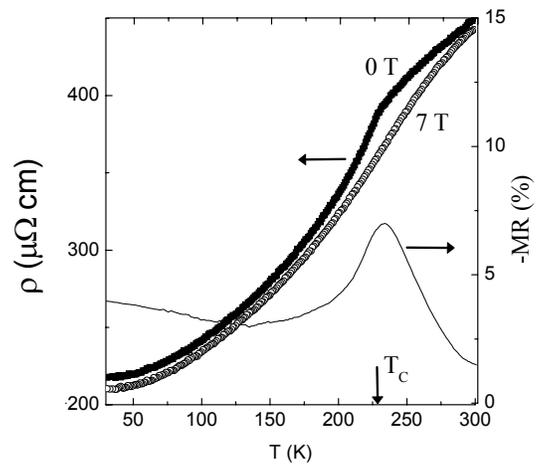

Fig. 6
Mahendiran et al.